\documentclass[11pt,a4paper]{article}

\usepackage[T1]{fontenc}
\usepackage[utf8]{inputenc}
\usepackage[english]{babel}
\usepackage[a4paper,margin=25mm]{geometry}
\usepackage{microtype}
\usepackage{setspace}
\usepackage{parskip}
\usepackage{titlesec}
\usepackage{fancyhdr}
\usepackage{authblk}
\usepackage{amsmath,amssymb,amsfonts,mathtools}
\usepackage{bm}
\usepackage{siunitx}
\usepackage{graphicx}
\usepackage{subcaption}
\usepackage{caption}
\usepackage{float}
\usepackage{booktabs}
\usepackage{array}
\usepackage{tabularx}
\usepackage{placeins}
\graphicspath{{figures/}}
\usepackage{xcolor}

\titleformat{\section}{\large\bfseries}{\thesection}{0.8em}{}
\titleformat{\subsection}{\normalsize\bfseries}{\thesubsection}{0.8em}{}

\title{Thermodynamically consistent initialization of the Maxwell--Cattaneo---Vernotte heat conduction model: Analytical solutions and engineering applications}
\author[1]{Zalán Sándor}
\author[1,2]{Róbert Kovács}
\date{\today}

\affil[1]{Department of Energy Engineering, Faculty of Mechanical Engineering, Budapest University of Technology and Economics, Műegyetem rkp. 3., H-1111 Budapest, Hungary}
\affil[2]{Department of Theoretical Physics, Wigner Research Centre for Physics, Institute for Particle and Nuclear Physics, Budapest, Hungary}

\begin{document}
\maketitle

\begin{abstract}
This paper investigates the numerical initialization of the one-dimensional Maxwell--Cattaneo--Vernotte (MCV) heat conduction model, addressing the critical mathematical challenge of defining the initial time derivative for non-equilibrium states. In modern high-frequency thermal engineering applications, traditional time-integration schemes and commercial finite-element solvers frequently introduce severe numerical artifacts when handling non-Fourier models. This study systematically evaluates three initialization strategies: a zero derivative, a spatially uniform non-zero derivative, and an exact space-dependent derivative. Using an explicit staggered finite-difference scheme, the transient responses to an exponentially distributed initial temperature field under adiabatic boundary conditions are compared against an analytical solution using the Galerkin method. The results demonstrate that assuming a zero or spatially uniform initial derivative introduces significant unphysical oscillatory deviations, leading to heat-flux prediction errors as the relaxation time increases. Conversely, mapping the exact space-dependent derivative onto the staggered grid preserves the local thermodynamic structure of the initial state, yielding robust transient responses that match the analytical benchmark without meaningful computational overhead. These findings establish a thermodynamically consistent initialization technique that can be extended to non-local models as well.
\end{abstract}

\section{Introduction}
\label{sec:intro}

Heat equations beyond Fourier have been known since the beginning of the 20th~century, initiated by the works of Tisza and Landau \cite{Tisza47, Lan47}. There are phenomena that point beyond diffusion, such as second sound, ballistic heat conduction, and the presence of multiple diffusion channels in heterogeneous materials \cite{McN74t, Kovacs2024PhysRep} and functionally graded materials \cite{Amiri25I, Amiri25II}. Among these, heat-wave models and finite-speed thermal propagation have also been reviewed in detail by Joseph and Preziosi \cite{JosephPreziosi1989}. The phenomenon of over-diffusion is particularly dominant in macroscale heterogeneous materials at room temperature, where parallel heat transfer channels interact without involving actual wave propagation \cite{Feher2025PhD}. Such deviations from Fourier's law are observed in low-temperature environments, nanostructures, and even in low-pressure states of fluids. In summary, the existence of multiple time scales in the transport process requires the use of an extended heat equation, regardless of whether such a model is hyperbolic or parabolic. 

Beyond classical cryogenic environments, the engineering relevance of non-Fourier heat conduction has expanded drastically in recent years. Modern practical applications such as ultra-fast laser processing of metals and dielectrics \cite{Malina16}, thermal management in 3D microelectronic packaging and integrated circuits \cite{Sharma20}, and metal additive manufacturing (e.g., selective laser melting) \cite{Gu21} rely on predicting extreme transient thermal responses. Furthermore, the thermal ablation of biological tissues using short-pulse lasers requires precise modeling of heat propagation to prevent unintended damage to surrounding healthy cells \cite{Walsh89, Jau08}. In these high-frequency, extreme thermal gradient environments, traditional parabolic models predict instantaneous heat conduction, which can lead to substantial errors in peak temperature and localized thermal stress predictions. In many of these situations, deviations from Fourier's law can be observed only during a transient process.
Therefore, one of the most relevant applications of non-Fourier equations can be the determination of thermal diffusivity (and thermal conductivity \cite{FehKov24}), 
typically using a heat pulse technique \cite{Parker1961}. This is a particularly outstanding method since one can adjust the pulse duration, influencing the time scale of the boundary condition. That boundary time scale must excite the corresponding heat transfer mechanisms in order to make the deviations observable. In the case of second sound, its possible occurrence in solids had already been discussed from a macroscopic heat-wave perspective by Chester \cite{Chester1963}. Later, the celebrated result of Guyer and Krumhansl \cite{GK64}, known as the window condition, helped researchers determine the proper excitation frequency, and the consistent observation of second sound became possible in numerous crystals. 

In most experimental techniques, the initial conditions describe a homogeneous equilibrium and thus require a zero initial time derivative for the homogeneous temperature distribution. The definition of the initial conditions of heat equations beyond Fourier is not straightforward when homogeneous equilibrium does not apply \cite{kovacs2022analytical}. Unlike the classical parabolic approach, a non-Fourier model features a constitutive relationship that is itself a partial differential equation restricting the time evolution of the heat current density. Consequently, solving it also requires the knowledge of the initial time derivatives of the field variables. Simply assuming these time derivatives to be zero -- a common trap when the thermodynamic origin is hidden by substituting the variables into a pure temperature representation -- can lead to seriously misleading assumptions \cite{kovacs2022analytical}. On the one hand, it influences the measurement outcome and the validity of the corresponding evaluation technique. On the other hand, non-zero initial time derivatives raise further questions about the solvability of heat equations beyond Fourier due to the various additional mathematical and physical options the evolution equations offer, which are highly relevant for the aforementioned advanced engineering tasks.

In the present work, we place our focus on the Maxwell--Cattaneo--Vernotte (MCV) equation (frequently briefly called the Cattaneo equation) \cite{Cattaneo1958, Vernotte1958} in which a memory-type extension is present, its one-dimensional constitutive equation reads
\begin{align}
\tau\,\partial_t q + q = -\lambda\,\partial_x T, \label{eq:mcv_article_dim_const}
\end{align}
where $T$ denotes the temperature, $q$ is the heat flux, $\lambda$ is the thermal conductivity and $\tau$ is called the relaxation time. In the present study, the model is used as the simplest non-Fourier extension, in which flux relaxation appears explicitly and requires an additional initial condition besides the initial temperature distribution. Furthermore, the MCV model is thermodynamically compatible, meaning it can be rigorously derived by exploiting the first and second laws of thermodynamics. Such compatibility ensures that the model exhibits asymptotically stable equilibrium solutions, keeping the predicted temperature fields physically admissible \cite{MullerRuggeri1998, VanFulop2012}. Although the practical engineering relevance of the MCV model is restricted to microscale or low-temperature heat conduction problems \cite{Auriault2016}, the growing applications of non-Fourier equations in modern engineering problems make it an excellent foundational example for demonstrating the necessity of careful model initialization. Moreover, the mathematical structure remains sufficiently clear for a systematic comparison between analytical and numerical results, without introducing any further complications regarding boundary conditions.

\begin{figure}[H]
    \centering
    \includegraphics[width=0.7\linewidth]{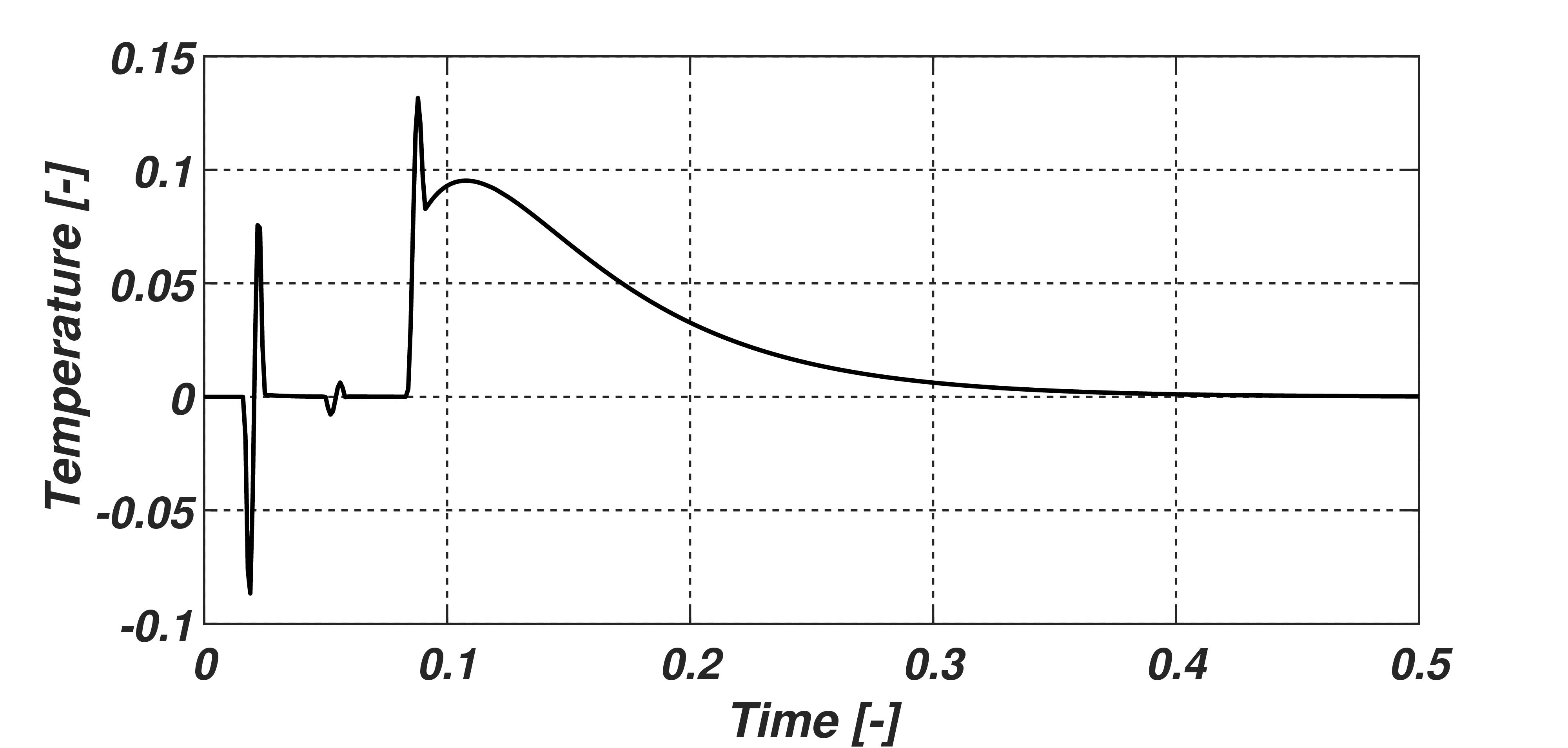}
    \caption{Demonstrating the appearance of artificial oscillations for hyperbolic heat equations in COMSOL \cite{RietEtal18}. It is difficult to distinguish between real physical wave propagation and nonphysical numerical error.}
    \label{fig:1}
\end{figure}

As an alternative to custom finite difference schemes, industrial finite element software such as COMSOL has been explored to solve generalized heat conduction models, such as the MCV equation \cite{RietEtal18}. However, implementing such non-Fourier models within commercial finite element environments is challenging and often computationally inefficient even for a one-dimensional situation. For instance, solving the Cattaneo equation in COMSOL using standard time-stepping methods, such as Runge--Kutta (RK34) or Backward Differentiation Formula (BDF), requires significant computational resources, leading to long run times and high memory demands compared to much simpler finite difference schemes \cite{RietEtal18}. Furthermore, these commercial solvers can easily introduce severe numerical artifacts, including dissipation errors that artificially reduce wave amplitudes and dispersion errors that introduce spurious, unphysical oscillations \cite{FulEtal20}. In certain cases, these artificial oscillations manifest independently of the chosen mesh or time step sizes. Consequently, it becomes exceedingly difficult to distinguish whether the observed wave-like phenomena are genuine physical characteristics of the non-equilibrium state or merely numerical errors introduced by the code itself. Due to these pronounced limitations and the difficulty of validating the output, relying solely on standard numerical packages to evaluate non-Fourier heat wave propagation is often inadequate. Figure \ref{fig:1} shows the emergence of artificial oscillations that significantly distort the obtained numerical solution, as reported in \cite{RietEtal18}.

While non-Fourier models are often rearranged as a single partial differential equation for temperature, this temperature representation is strictly valid only for linear models with constant coefficients. Retaining both the temperature and the heat flux as primary field variables -- rather than eliminating either the heat flux or the temperature -- provides a much more flexible and reliable framework for implementing complex initial and boundary conditions. In the following, we consider only rigid isotropic materials with linear transport coefficients, thus $\lambda$ and $\tau$ are constant scalars. Furthermore, Eq.~\eqref{eq:mcv_article_dim_const} is coupled with the balance of internal energy, in a one-dimensional configuration without heat sources,
\begin{align}
\rho c\,\partial_t T + \partial_x q = 0, \label{eq:mcv_article_dim_bal}
\end{align}
in which $c$ is the heat capacity and $\rho$ is the density. Although the following is beyond the scope of our paper, we note that space and time-dependent heat sources can significantly modify the dynamics and can result in a different equilibrium than that of Fourier's law \cite{Kovacs2024PhysRep}. We want to focus our study on the consequences of initial conditions, using a non-homogeneous initial temperature distribution. We also note that besides the $T$-representation of a heat equation, its $q$-representation exists as well \cite{Kovacs2024PhysRep}. Consequently, the temperature as a state variable loses its meaning, and thus cannot be used either as an initial condition or boundary condition in a $q$-representation. Moreover, in such a case, the initial heat flux field and its time derivative would offer the necessary set of initial data, but since none of them is practically measurable, we do not study this situation in detail and restrict ourselves to the investigation of the complete heat equation as a system of partial differential equations.

The scientific novelty of this work lies in the rigorous decoupling of non-Fourier physical phenomena from numerical initialization artifacts. While existing literature extensively investigates the formulation of extended heat equations and their steady-state or long-term transient behavior, the precise mathematical treatment of non-equilibrium initial states remains a significant gap. By systematically analyzing the initial constitutive response—specifically, the discrete realization of the initial time derivative—we establish a mathematically consistent framework that bridges thermodynamic theory and numerical implementation. In order to avoid any numerical artifacts, our analysis is based on analytical solutions using Galerkin's technique \cite{kovacs2022analytical} (for more advanced applications, we refer to \cite{Huang26}). Additionally, we use a finite difference technique on a staggered grid to demonstrate the numerical implementation of the non-homogeneous initial states, even when prescribing nonzero initial time derivatives.
Ultimately, we demonstrate how different mathematical initializations fundamentally alter the predicted early-time thermal response, ensuring that computational predictions in advanced applications reflect actual material behavior rather than arbitrary numerical approximations.

\section{The non-equilibrium initial-boundary value problem}
\label{sec:mcv_ibc_article}

In order to systematically investigate the consequences of non-equilibrium initial states, we must define a complete initial-boundary value problem based on the coupled system of the energy balance \eqref{eq:mcv_article_dim_bal} and the MCV constitutive equation \eqref{eq:mcv_article_dim_const}. As established in the introduction, we avoid eliminating any of the variables to form a single higher-order partial differential equation. Instead, we keep both the temperature $T(t,x)$ and the heat flux $q(t,x)$ as primary field variables. This coupled approach is essential to maintain thermodynamic consistency. It keeps the relationship between the initial conditions and the constitutive equation transparent, ensuring that the physical compatibility of the initial time derivatives is not inadvertently violated during the solution process.

The investigated problem is a one-dimensional transient heat-conduction process on a finite interval, recalling the governing equations,
\begin{align}
\rho c\,\partial_t T + \partial_x q &= 0, \\
\tau\,\partial_t q + q &= -\lambda\,\partial_x T. 
\end{align}
The initial temperature field is prescribed in exponential form,
\begin{align}
T(0,x) = T_\mathrm{ref}\exp\left(-\frac{x}{z}\right),
\end{align}
where $T_\mathrm{ref}$ is a characteristic temperature scale and $z$ can be used to control the steepness of the initial profile, following \cite{kovacs2022analytical}. This choice provides a smooth, spatially heterogeneous initial condition, which is well-suited to demonstrating why a systematic investigation is necessary in the case of non-Fourier heat conduction models. Such an initial profile may arise, for example, in flash experiments performed on semitransparent materials, where the heat pulse is absorbed not only at the front surface but also within the body, characteristic of semi-transparent objects such as biological tissues, polymers, or glasses \cite{ZolfMaer11b}. Physically, this exponential profile closely approximates the volumetric energy source dictated by the Beer--Lambert law, where $z$ acts as the effective optical penetration depth of the material. By initiating the heat conduction process immediately after this fast energy absorption, we effectively decouple the heating phase and initialize the system in a strictly thermal, non-equilibrium state. Furthermore, from a mathematical perspective, the exponential function is highly advantageous. Unlike a step function or a rectangular spatial pulse -- which can introduce unphysical singularities or severe numerical artifacts such as the Gibbs phenomenon -- the smooth nature of the exponential profile ensures that all spatial derivatives remain well-behaved in the analytical solution, and does not distort the Fourier series expansion. This regularity is particularly beneficial when testing numerical schemes or analytical solution techniques, as it allows for a clean separation of genuine non-Fourier physical effects from purely numerical anomalies. The central question of this problem is the determination of the initial heat flux field and the corresponding initial time derivatives.

At the two ends of the domain, adiabatic boundary conditions are imposed,
\begin{align}
q(t,0)=0, \qquad q(t,L)=0,
\end{align}
in order to keep the dynamics isolated and free from any further effects. For example, if the temperature or convection boundary conditions were defined, then these can stabilize the dynamics, introducing additional (physical) dissipation to the system, but we are interested in the pure dynamical behavior of the MCV equation when various initialization strategies are applied. 
From the physical point of view, these conditions are simple, yet they are already sufficient to reveal the main difficulty of the MCV model. In a Fourier-type problem, the initial temperature field is enough to start the process. In the present case, however, the heat flux must satisfy its own evolution equation. Consequently, the initial state cannot be completely characterized by $T(0,x)$ and $q(0,x)$ alone when executing a numerical time step. Some additional assumptions are needed regarding the initial constitutive response, and this is where different numerical implementations begin to diverge. In the present paper, this issue is treated through the initial time derivative of the heat flux, which becomes the central object of comparison in the later sections.

\section{Applying the Galerkin technique} \label{sec:mcv_galerkin_article}

For the analysis, it is convenient to introduce a dimensionless form of the governing equations. The dimensionless temperature, spatial coordinate, and time coordinate are chosen as
\begin{equation}\label{eq:mcv_article_dimless_basic}
    \vartheta := \frac{T}{T_\mathrm{ref}},
    \qquad
    \xi := \frac{x}{L},
    \qquad
    \mathrm{Fo} := \frac{a t}{L^2},
\end{equation}
where $a=\lambda/(\rho c)$ is the thermal diffusivity. These three scales are prescribed directly: $T_\mathrm{ref}$ is the characteristic temperature scale of the initial condition, $L$ is the length of the investigated interval, and $L^2/a$ is the diffusive time scale. Once these choices have been made, the scaling of the heat flux and the relaxation time follows from the dimensional MCV system. The corresponding dimensionless quantities are
\begin{equation}\label{eq:mcv_article_dimless_flux_tau}
    \hat{q} = \frac{L}{\lambda T_\mathrm{ref}}\,q,
    \qquad
    \hat{\tau} = \frac{a\tau}{L^2}.
\end{equation}
With these definitions, the MCV system takes the form
\begin{align}
\hat{\tau}\,\partial_{\mathrm{Fo}}\hat{q}(\mathrm{Fo},\xi)+\hat{q}(\mathrm{Fo},\xi) &= -\partial_{\xi}\vartheta(\mathrm{Fo},\xi)\,, \label{eq:mcv_article_dimless_const}\\
\partial_{\mathrm{Fo}}\vartheta(\mathrm{Fo},\xi)+\partial_{\xi}\hat{q}(\mathrm{Fo},\xi) &= 0\,. \label{eq:mcv_article_dimless_bal}
\end{align}
In this form, the deviation from the Fourier limit is governed by the single parameter $\hat{\tau}$, which makes the comparison between different transient regimes particularly clear.

Galerkin's method is utilized here in two ways: it provides the exact analytical reference solution (validation) for the numerical solution, and it mathematically formalizes the implementation of the initial conditions. Within this framework, both fields are represented as a product of time-dependent coefficients and space-dependent basis functions,
\begin{align}
\hat{q}(\mathrm{Fo},\xi) &\approx \sum_{n=1}^{N} a_n(\mathrm{Fo})\,\sin(n\pi\xi), \label{eq:mcv_article_gal_q}\\
\vartheta(\mathrm{Fo},\xi) &\approx b_{00} + \sum_{n=1}^{N} b_n(\mathrm{Fo})\,\cos(n\pi\xi), \label{eq:mcv_article_gal_th}
\end{align}
where $b_{00}$ represents the steady-state mean temperature, and the selected trigonometric basis functions identically satisfy the adiabatic boundary conditions. After substituting these series into the governing equations \eqref{eq:mcv_article_dimless_const}--\eqref{eq:mcv_article_dimless_bal} and utilizing the orthogonality of the basis functions, the approximation yields a decoupled equation for the zero-mode ($b_{0}'(\mathrm{Fo})=0 \implies b_0(\mathrm{Fo}) = b_{00}$) and a coupled linear system of ordinary differential equations for each mode $k \ge 1$:
\begin{align}
\hat{\tau}\,a_k'(\mathrm{Fo})+a_k(\mathrm{Fo})-k\pi\,b_k(\mathrm{Fo}) &= 0, \label{eq:mcv_article_mode_1}\\
b_k'(\mathrm{Fo})+k\pi\,a_k(\mathrm{Fo}) &= 0. \label{eq:mcv_article_mode_2}
\end{align}

\subsection{Eigenvalue problem and general solution}
To determine the analytical solution, the system of ordinary differential equations can be rearranged into a standard matrix form,
\begin{equation}
\begin{bmatrix} a_k'(\mathrm{Fo}) \\ b_k'(\mathrm{Fo}) \end{bmatrix} = \begin{bmatrix} -\frac{1}{\hat{\tau}} & \frac{k\pi}{\hat{\tau}} \\ -k\pi & 0 \end{bmatrix} \begin{bmatrix} a_k(\mathrm{Fo}) \\ b_k(\mathrm{Fo}) \end{bmatrix}.
\end{equation}
The characteristic equation of the coefficient matrix is $s^2 + \frac{1}{\hat{\tau}}s + \frac{(k\pi)^2}{\hat{\tau}} = 0$, which yields the fundamental eigenvalues:
\begin{equation}
s_{1,k} = \frac{-1 + \sqrt{1 - 4\hat{\tau}(k\pi)^2}}{2\hat{\tau}}, \qquad s_{2,k} = \frac{-1 - \sqrt{1 - 4\hat{\tau}(k\pi)^2}}{2\hat{\tau}}.
\end{equation}
The corresponding eigenvectors, defining the phase-space direction of the thermal modes, are
\begin{equation}
\mathbf{v}_{1,k} = \begin{bmatrix} 1 \\ v_{12,k} \end{bmatrix} = \begin{bmatrix} 1 \\ \frac{\hat{\tau}}{k\pi}\left(\frac{1}{\hat{\tau}} + s_{1,k}\right) \end{bmatrix}, \qquad \mathbf{v}_{2,k} = \begin{bmatrix} 1 \\ v_{22,k} \end{bmatrix} = \begin{bmatrix} 1 \\ \frac{\hat{\tau}}{k\pi}\left(\frac{1}{\hat{\tau}} + s_{2,k}\right) \end{bmatrix}.
\end{equation}
Provided that the discriminant is non-zero ($s_{1,k} \neq s_{2,k}$), the general solution for each mode is a linear combination of these states:
\begin{equation}\label{eq:mcv_general_solution}
\begin{bmatrix} a_k(\mathrm{Fo}) \\ b_k(\mathrm{Fo}) \end{bmatrix} = C_{1,k} e^{s_{1,k} \mathrm{Fo}} \begin{bmatrix} 1 \\ v_{12,k} \end{bmatrix} + C_{2,k} e^{s_{2,k} \mathrm{Fo}} \begin{bmatrix} 1 \\ v_{22,k} \end{bmatrix}.
\end{equation}

\subsection{Critical relaxation time and resonance}
The analytical method reveals a physical transition point where the eigenvalues are resonant. This occurs when the discriminant vanishes, defining a critical relaxation time for each mode $k$,
\begin{equation}
\hat{\tau}^* = \frac{1}{4(k\pi)^2}.
\end{equation}
At this critical value, the standard general solution fails because only a single independent eigenvector exists ($\mathbf{v}_k = [1, \frac{1}{2k\pi}]^\top$) corresponding to the repeated eigenvalue $s_k = -2(k\pi)^2$. To maintain mathematical rigor and avoid singularities during the investigation, the solution at $\hat{\tau}^*$ must be obtained using the generalized eigenvector $\mathbf{u}_k = [-\frac{1}{2(k\pi)^2}, 0]^\top$:
\begin{equation}\label{eq:mcv_critical_solution}
\begin{bmatrix} a_k^*(\mathrm{Fo}) \\ b_k^*(\mathrm{Fo}) \end{bmatrix} = e^{s_k \mathrm{Fo}} \left( C_{1,k}^* \mathbf{v}_k + C_{2,k}^* (\mathrm{Fo}\, \mathbf{v}_k + \mathbf{u}_k) \right).
\end{equation}

\subsection{Implementing the initial conditions}
The constants of integration ($C_{1,k}$ and $C_{2,k}$) are uniquely determined by the initial macroscopic state. For the exponential initial temperature profile, the initial thermal modal coefficients $b_k(0) = b_{0k}$ are obtained via standard Fourier expansion, that is,
\begin{equation}
b_{0k} = \frac{2(L/z)\big(1-(-1)^k \exp(-L/z)\big)}{(L/z)^2 + (k\pi)^2}, \qquad k \ge 1.
\end{equation}
The second set of constraints arises from the chosen initial dynamic state of the system, fundamentally differentiating the initializations.

\textbf{Case 1: Zero initial temperature derivative.} Imposing $\left.\partial_{\mathrm{Fo}}\vartheta\right|_{\mathrm{Fo}=0}=0$ implies $b_k'(0) = 0$. Applying this to Eq.~\eqref{eq:mcv_general_solution} yields:
\begin{equation}
C_{1,k} = \frac{b_{0k}}{v_{12,k}\left(1 - \frac{s_{1,k}}{s_{2,k}}\right)}, \qquad C_{2,k} = \frac{b_{0k}}{v_{22,k}\left(1 - \frac{s_{2,k}}{s_{1,k}}\right)}.
\end{equation}
If the system is at the critical relaxation time ($\hat{\tau}^*$), applying this to Eq.~\eqref{eq:mcv_critical_solution} yields $C_{1,k}^* = 2k\pi b_{0k}$ and $C_{2,k}^* = 4(k\pi)^3 b_{0k}$.

\textbf{Case 2: Zero initial heat-flux derivative.} Conversely, imposing $\left.\partial_{\mathrm{Fo}}\hat{q}\right|_{\mathrm{Fo}=0}=0$ implies $a_k'(0) = 0$, producing a different set of coefficients:
\begin{equation}
C_{1,k} = \frac{b_{0k}}{v_{12,k} - v_{22,k}\frac{s_{1,k}}{s_{2,k}}}, \qquad C_{2,k} = \frac{b_{0k}}{v_{22,k} - v_{12,k}\frac{s_{2,k}}{s_{1,k}}}.
\end{equation}
At the critical relaxation time, this assumption yields $C_{1,k}^* = 2k\pi b_{0k}$ and $C_{2,k}^* = 2(k\pi)^3 b_{0k}$.

\begin{figure}[H]
    \centering
    \includegraphics[width=0.8\linewidth]{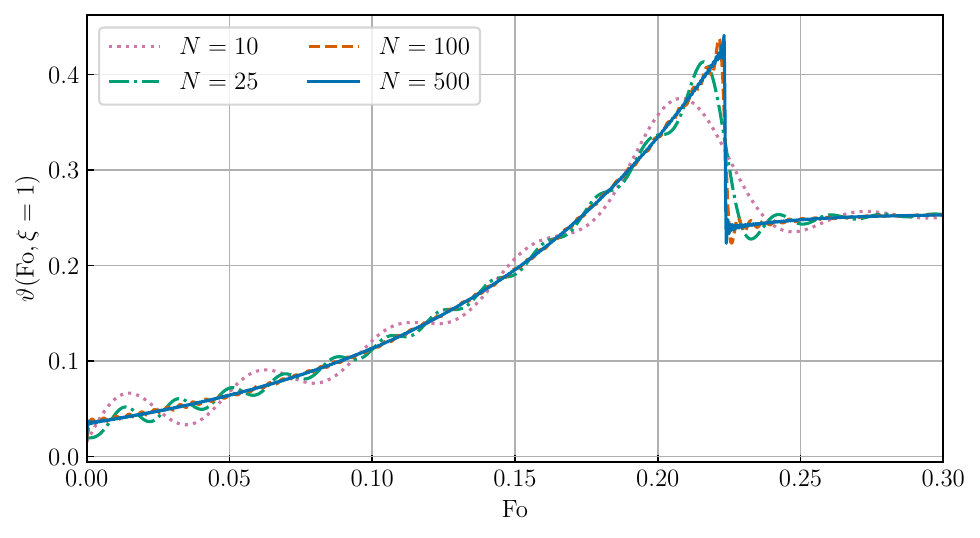}\\
    \includegraphics[width=0.8\linewidth]{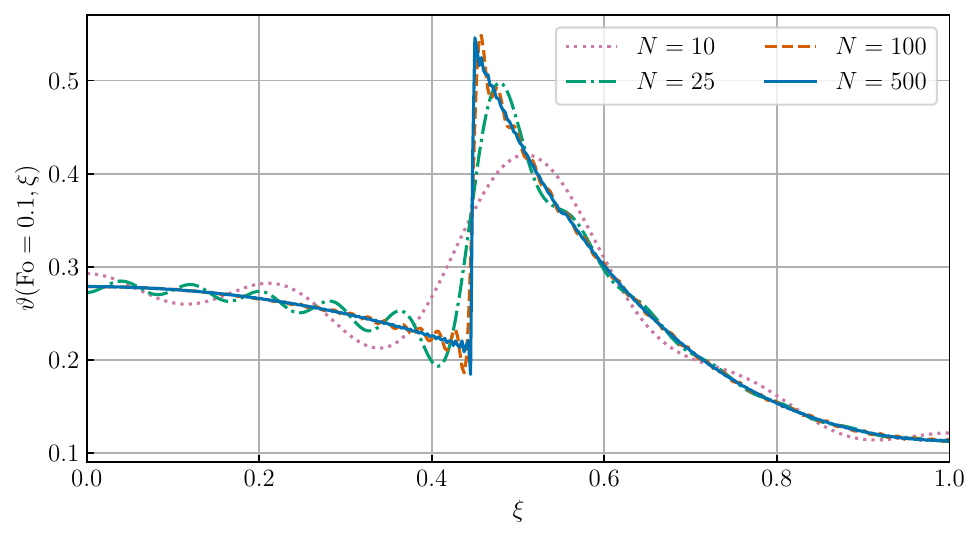}
    \caption{Convergence test of the Galerkin-type analytical solution in the case when the initial time derivative of $q$ is zero and \(\hat \tau = 0.05\). Top figure: convergence of the temperature in time. Bottom figure: convergence of the temperature in space. Such a study makes clear that the proper representation of sharp fronts requires the use of 500 terms of the Galerkin series.}
    \label{fig:convergence}
\end{figure}

\textbf{Case 3: Arbitrary initial heat flux.} The most generalized non-equilibrium initialization does not restrict either time derivative to zero. Instead, it allows for the prescription of an arbitrary, independent initial heat flux field, represented by the modal coefficients $a_k(0) = a_{0k}$. In this fully dynamic state, the time derivatives are given by the initial values of the state variables via the governing balance and constitutive equations. Using the general solution \eqref{eq:mcv_general_solution} evaluated at $\mathrm{Fo}=0$, this condition yields the algebraic system,
\begin{equation}
C_{1,k} + C_{2,k} = a_{0k}, \qquad C_{1,k} v_{12,k} + C_{2,k} v_{22,k} = b_{0k}.
\end{equation}
Solving this system provides the exact integration constants for an arbitrary initial state:
\begin{equation}
C_{1,k} = \frac{b_{0k} - a_{0k} v_{22,k}}{v_{12,k} - v_{22,k}}, \qquad C_{2,k} = \frac{a_{0k} v_{12,k} - b_{0k}}{v_{12,k} - v_{22,k}}.
\end{equation}
If the system operates exactly at the critical relaxation time ($\hat{\tau}^*$), applying this fully generalized initial condition to Eq.~\eqref{eq:mcv_critical_solution} yields $C_{1,k}^* = 2k\pi b_{0k}$ and $C_{2,k}^* = 4(k\pi)^3 b_{0k} - 2(k\pi)^2 a_{0k}$. By implementing these coefficients, the analytical reference can handle any combination of initial temperature and heat-flux fields, automatically capturing the correct non-zero initial time derivatives dictated by the coupled model. However, in the present paper, we do not aim to further study this situation since this is based on an arbitrary heat flux profile, leaving the outcomes inconclusive for particular applications.

\section{Basis of the finite difference scheme}\label{sec:mcv_dimless_num_article}

The numerical method is an explicit finite-difference scheme applied directly to the coupled system. The spatial interval $\xi\in[0,1]$ is divided into $N_x$ cells of equal length $h_\xi=1/N_x$, while the time interval is divided into $N_t$ steps with step size $h_{\mathrm{Fo}}$. The lower index denotes the time level, therefore $j=0$ corresponds to the initial state and $j+1$ to the next time level. The upper index denotes the spatial index of the corresponding numerical vector. The temperature vector contains $N_x$ unknowns, indexed by $i=0,1,\ldots,N_x-1$, while the heat-flux vector contains $N_x+1$ values, indexed by $i=0,1,\ldots,N_x$, because the two boundary faces are also included. 

\begin{figure}[!htb]
    \centering
    \includegraphics[width=0.75\linewidth]{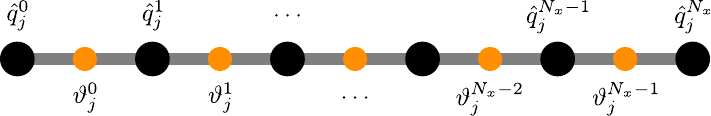}
    \caption{Schematic representation of the spatial discretization used in the explicit finite-difference simulations. The temperature is evaluated at cell centers, while the heat flux is evaluated at cell faces.}
    \label{fig:mcv_article_disc}
\end{figure}

As shown in Figure \ref{fig:mcv_article_disc}, the temperature values are assigned to cell centers, and the heat-flux values are assigned to cell faces. The half-cell shift, however, is not introduced as a separate indexing convention in the numerical vectors. It only has to be taken into account when the continuous initial temperature profile is sampled at the cell centers.

The discrete and yet not complete set of initial and boundary conditions is written as
\begin{align}
    & \vartheta_0^{\,i} = \exp\!\left(-\frac{L}{z}\,\left(i+\frac{1}{2}\right)h_\xi\right), && i\in[0,\,N_x-1], \label{eq:mcv_article_init_disc_theta}\\
    & \hat{q}_j^{\,0}=0, && j\in[0,\,N_t], \label{eq:mcv_article_left_bc}\\
    & \hat{q}_j^{\,N_x}=0, && j\in[0,\,N_t]. \label{eq:mcv_article_right_bc}
\end{align}
The additional MCV-specific time derivative term appears in the very first time step. Instead of starting directly from the standard update, the first interior flux field is initialized according to
\begin{equation}\label{eq:mcv_article_init_q}
    \hat{q}_{0}^{\,i+1}
    =
    -\frac{1}{h_\xi}\,\Big(\vartheta_{0}^{\,i+1}-\vartheta_{0}^{\,i}\Big)
    -\hat{\tau}\,\Delta_{\mathrm{Fo}}\hat{q}_{0}^{\,i+1},
    \qquad i\in[0,\,N_x-2].
\end{equation}
Here $\hat{q}_{0}^{\,i+1}$ is the heat flux at the interior face between the neighbouring temperature cells $i$ and $i+1$. The same spatial index is used for $\Delta_{\mathrm{Fo}}\hat{q}_{0}^{\,i+1}$, because this term belongs to the same interior heat-flux component. The notation $\Delta_{\mathrm{Fo}}\hat{q}_{0}^{\,i}$ is used deliberately instead of the continuous derivative symbol $\partial_{\mathrm{Fo}}\hat{q}$. It denotes the value inserted into the finite-difference scheme as the numerical representation of the initial time derivative of the dimensionless heat flux. This is the term whose interpretation distinguishes the three numerical cases studied later.

For the subsequent time levels, the explicit scheme follows from the discretization of Eqs.~\eqref{eq:mcv_article_dimless_const}--\eqref{eq:mcv_article_dimless_bal}. The resulting iteration formulas are
\begin{align}
    \hat{q}_{j+1}^{\,i+1}
    &=
    \left(1-\frac{h_{\mathrm{Fo}}}{\hat{\tau}}\right)\hat{q}_{j}^{\,i+1}
    -\frac{h_{\mathrm{Fo}}}{\hat{\tau}h_\xi}\,\Big(\vartheta_j^{\,i+1}-\vartheta_j^{\,i}\Big),
    && j\in[0,\,N_t-1],\ i\in[0,\,N_x-2],
    \label{eq:mcv_article_iter_q}\\
    \vartheta_{j+1}^{\,i}
    &=
    \vartheta_j^{\,i}
    -\frac{h_{\mathrm{Fo}}}{h_\xi}\,\Big(\hat{q}_{j+1}^{\,i+1}-\hat{q}_{j+1}^{\,i}\Big),
    && j\in[0,\,N_t-1],\ i\in[0,\,N_x-1].
    \label{eq:mcv_article_iter_th}
\end{align}
These formulas are deliberately simple. The main aim of the study was not to optimize the time integrator itself, but to investigate how strongly the initialization affects the quality of the solution when the interior scheme is kept fixed. However, because the scheme is fully explicit, the chosen step sizes $h_{\mathrm{Fo}}$ and $h_\xi$ cannot be entirely arbitrary. { The hyperbolic nature of the MCV equation implies a finite propagation speed of thermal waves, corresponding to a dimensionless speed of $1/\sqrt{\hat{\tau}}$ \cite{OzisikTzou1994}.} Consequently, the discretization must strictly satisfy a stability condition to prevent numerical blow-up, thereby restricting the allowable time step more severely than pure parabolic diffusion models.

\subsection{Von Neumann stability analysis} \label{sec:stab_analysis}
To derive the exact stability criterion, we apply the von Neumann method by introducing discrete Fourier modes on the staggered grid. Let the discrete fields be represented as
\begin{equation}
\hat{q}_j^i = Q^j e^{\imath m i h_\xi}, \qquad \vartheta_j^i = \Theta^j e^{\imath m (i-0.5) h_\xi},
\end{equation}
where $\imath$ is the imaginary unit and $m$ is the spatial wave number. Substituting these into the staggered explicit iteration formulas \eqref{eq:mcv_article_iter_q}--\eqref{eq:mcv_article_iter_th} yields the amplification matrix $\mathbf{G}$, defined by $[Q^{j+1}, \Theta^{j+1}]^\top = \mathbf{G} [Q^j, \Theta^j]^\top$,
\begin{equation}
\mathbf{G} = \begin{bmatrix} 
1 - \frac{h_{\mathrm{Fo}}}{\hat{\tau}} & -\imath \frac{2h_{\mathrm{Fo}}}{\hat{\tau}h_\xi}\sin(k) \\ 
-\imath \frac{2h_{\mathrm{Fo}}}{h_\xi}\left(1 - \frac{h_{\mathrm{Fo}}}{\hat{\tau}}\right)\sin(k) & 1 - \frac{4h_{\mathrm{Fo}}^2}{\hat{\tau}h_\xi^2}\sin^2(k) 
\end{bmatrix},
\end{equation}
where $k = m h_\xi / 2$. The characteristic polynomial of the amplification matrix is $P(\lambda) = \lambda^2 - \mathrm{Tr}(\mathbf{G})\lambda + \mathrm{Det}(\mathbf{G}) = 0$. The invariants are:
\begin{equation}
\mathrm{Tr}(\mathbf{G}) = 2 - \frac{h_{\mathrm{Fo}}}{\hat{\tau}} - \frac{4h_{\mathrm{Fo}}^2}{\hat{\tau}h_\xi^2}\sin^2(k), \qquad \mathrm{Det}(\mathbf{G}) = 1 - \frac{h_{\mathrm{Fo}}}{\hat{\tau}}.
\end{equation}
According to the Jury stability criteria (also applied for various non-Fourier models in \cite{RietEtal18}), the roots of the characteristic polynomial lie within the unit circle if and only if $1 - \mathrm{Tr}(\mathbf{G}) + \mathrm{Det}(\mathbf{G}) \ge 0$, $1 + \mathrm{Tr}(\mathbf{G}) + \mathrm{Det}(\mathbf{G}) \ge 0$, and $\mathrm{Det}(\mathbf{G}) \le 1$. The first and third conditions are inherently satisfied for physically (and, thus, thermodynamically) meaningful (positive) transport parameter sets. The second condition evaluates to
\begin{equation}
4 - 2\frac{h_{\mathrm{Fo}}}{\hat{\tau}} - \frac{4h_{\mathrm{Fo}}^2}{\hat{\tau}h_\xi^2}\sin^2(k) \ge 0.
\end{equation}
In order to guarantee stability for all possible wave numbers, we evaluate the most restrictive case ($\sin^2(k) = 1$), yielding the exact strict stability constraint for the staggered MCV scheme,
\begin{equation}\label{eq:mcv_exact_stability}
2\hat{\tau} - h_{\mathrm{Fo}} - 2\frac{h_{\mathrm{Fo}}^2}{h_\xi^2} \ge 0 \implies h_{\mathrm{Fo}} \le \frac{h_\xi^2}{4} \left( \sqrt{1 + \frac{16\hat{\tau}}{h_\xi^2}} - 1 \right).
\end{equation}
For fine spatial discretizations ($h_\xi \ll 1$), the quadratic term inside the root dominates, and the constraint simplifies asymptotically to the hyperbolic Courant-Friedrichs-Lewy (CFL) condition,
\begin{equation}\label{eq:mcv_cfl_limit}
h_{\mathrm{Fo}} \le h_\xi \sqrt{\hat{\tau}}.
\end{equation}
This analysis mathematically proves that the allowable time step is restricted by the finite propagation speed of thermal waves ($c = 1/\sqrt{\hat{\tau}}$), rather than pure diffusion.

\section{Numerical implementation of the initial derivative}\label{sec:mcv_init_deriv_article}

With the stability limits defined, we can systematically compare three discrete implementations of the initialization term $\Delta_{\mathrm{Fo}}\hat{q}_{0}^{i}$ in Eq.~\eqref{eq:mcv_article_init_q}. The central contribution of the paper is the comparison of three ways of interpreting the term $\Delta_{\mathrm{Fo}}\hat{q}_{0}^{i}$ in Eq.~\eqref{eq:mcv_article_init_q}. All three variants start from the same initial temperature field, the same adiabatic boundaries, and the same interior explicit iteration. The difference lies only in the way the initial constitutive response is introduced into the very first time step.

\subsection{Zero initial time derivative}\label{subsec:mcv_zero_deriv_article}
The mathematically simplest initialization is
\begin{equation}
\Delta_{\mathrm{Fo}}\hat{q}_{0}^{\,i+1}=0, \qquad i\in[0,\,N_x-2]\,.
\end{equation}
This assumption implicitly initializes the discrete system as if it were a purely Fourier process during the first time step. While computationally attractive because it requires no auxiliary calculations, it fundamentally contradicts the thermodynamics implied by the MCV equation unless the initial state is a homogeneous equilibrium. As the later results show, this approximation can still work well when $\hat{\tau}$ is very small, but the error grows rapidly once the non-Fourier character of the model becomes more pronounced. Physically, this implicitly assumes that the initial heat flux is stationary, completely ignoring the dynamic relaxation forced by the non-equilibrium temperature gradient.

\subsection{Spatially uniform non-zero derivative}\label{subsec:mcv_const_deriv_article}
A natural subsequent step is to replace the zero heat flux derivative with a spatially uniform non-zero constant. Rather than prescribing the initial time derivative of the heat flux arbitrarily, its value is derived by assuming that the initial time derivative of the temperature field vanishes, i.e., $\left.\partial_{\mathrm{Fo}}\vartheta\right|_{\mathrm{Fo}=0}=0$.

To evaluate this condition, the exponential initial temperature profile is expanded in the cosine basis as
\begin{equation}\label{eq:mcv_article_initial_cos_series}
    \vartheta_0(\xi)
    = b_{00}+\sum_{n=1}^{N} b_{n0}\cos(n\pi\xi),
\end{equation}
where the relevant modal Fourier coefficients are determined by
\begin{equation}\label{eq:mcv_article_bn0}
    b_{n0}
    =2\int_0^1 \exp\!\left(-\frac{L}{z}\xi\right)\cos(n\pi\xi)\,\mathrm{d}\xi
    =\frac{2(L/z)\big(1-(-1)^n\exp(-L/z)\big)}{(L/z)^2+(n\pi)^2},
    \qquad n\geq 1,
\end{equation}
with $b_{n0} := b_n(\mathrm{Fo}=0)$. The assumption $\left.\partial_{\mathrm{Fo}}\vartheta\right|_{\mathrm{Fo}=0}=0$ implies $b_n'(0)=0$. Utilizing the modal balance equation \eqref{eq:mcv_article_mode_2}, this requirement dictates that
\begin{equation}\label{eq:mcv_article_an0_from_temperature_derivative}
    b_n'(0)+n\pi a_n(0)=0
    \qquad\Longrightarrow\qquad
    a_n(0)=0.
\end{equation}
Substituting $a_n(0)=0$ into the modal constitutive equation \eqref{eq:mcv_article_mode_1} at $\mathrm{Fo}=0$ yields the initial derivative of the heat-flux modal coefficients,
\begin{equation}\label{eq:mcv_article_an0_derivative}
    a_n'(0)=\frac{n\pi}{\hat{\tau}}\,b_{n0}.
\end{equation}
Differentiating the heat-flux expansion \eqref{eq:mcv_article_gal_q} with respect to $\mathrm{Fo}$ and substituting $a_n'(0)$ yields the continuous, spatially dependent initial derivative field,
\begin{equation}\label{eq:mcv_article_qderiv_field}
\left.\partial_{\mathrm{Fo}}\hat{q}(\mathrm{Fo},\xi)\right|_{\mathrm{Fo}=0}
=
\sum_{n=1}^{N}
\frac{n\pi}{\hat{\tau}}
\frac{2(L/z)\big(1-(-1)^n\exp(-L/z)\big)}{(L/z)^2+(n\pi)^2}
\sin(n\pi\xi).
\end{equation}
The spatially uniform approximation is then obtained by averaging this field over the spatial domain:
\begin{equation}\label{eq:mcv_article_qderiv_const}
\Delta_{\mathrm{Fo}}\hat{q}_{0}
:=
\int_0^1
\left.\partial_{\mathrm{Fo}}\hat{q}(\mathrm{Fo},\xi)\right|_{\mathrm{Fo}=0}
\,\mathrm{d}\xi \,.
\end{equation}
The discrete initialization term in Eq.~\eqref{eq:mcv_article_init_q} is then assigned as $\Delta_{\mathrm{Fo}}\hat{q}_{0}^{\,i+1}=\Delta_{\mathrm{Fo}}\hat{q}_{0}$ for all interior faces $i\in[0,\,N_x-2]$. Although this approach retains low computational overhead, compressing a spatially varying dynamic response into a single scalar average introduces a step-like discontinuity relative to the constant boundary conditions, thereby compromising transient accuracy and introducing severe non-physical oscillations.

\subsection{Space-dependent derivative}
\label{subsec:mcv_field_deriv_article}
The most challenging initialization utilizes the exact, space-dependent, analytically derived field directly on the finite-difference grid. This method strictly preserves the local thermodynamic structure of the initial non-equilibrium state, hence,
\begin{equation}\label{eq:mcv_article_qderiv_pointwise}
\Delta_{\mathrm{Fo}}\hat{q}_{0}^{\,i+1}
:=
\sum_{n=1}^{N}
\frac{n\pi}{\hat{\tau}}
\frac{2(L/z)\big(1-(-1)^n\exp(-L/z)\big)}{(L/z)^2+(n\pi)^2}
\sin\!\left(n\pi \left(i+1\right) h_\xi\right),
\qquad i\in[0,\,N_x-2].
\end{equation}
By initializing the explicit scheme with the exact local derivatives mapped to the staggered grid, we can separate non-Fourier physical phenomena from unwanted numerical initialization artifacts. The additional computational cost may already be significant even in the present one-dimensional setting, since the summation up to \(N\) must be evaluated for all \(N_x-1\) interior heat-flux components. At the same time, this higher computational effort yields a highly accurate representation of the transient response. However, the computational overhead is strictly localized to the initialization phase. Furthermore, because the chosen exponential temperature profile is smooth, the Fourier coefficients $b_{n0}$ decay proportionally to $\mathcal{O}(1/n^2)$. This ensures rapid spectral convergence, allowing the series to be safely truncated at a moderate $N$ without introducing significant aliasing errors.

\section{Results and comparison}\label{sec:mcv_results_article}
The numerical experiments were carried out for three representative values of the dimensionless relaxation time, $\hat{\tau}\in\{0.001,\,0.01,\,0.05\}$. This range spans a nearly diffusive regime, an intermediate case, and a distinctly non-Fourier wave response. Since the study is based on an analytical solution, any additional relaxation time values can be used without restriction. Naturally, the numerical implementation may require very small time steps to keep the scheme stable. In all runs, the spatial domain was discretized into $N_x = 100$ cells. The dimensionless time step $h_{\mathrm{Fo}}$ was chosen to satisfy the stability limit derived in Section~\ref{sec:stab_analysis}, ensuring that no numerical blow-up occurred. The only modified ingredient across the simulations was the discrete initialization strategy. Moreover, the present numerical scheme is already validated across various heat equations \cite{RietEtal18}.

The comparison is organized around the three initialization strategies, because this is the point where the qualitative behaviour of the simulations begins to differ. The figures below show the numerical and analytical temperature and heat-flux histories for the selected values of $\hat{\tau}$. The purpose of the discussion is not only to identify the smallest error, but also to clarify how the different implementations affect the early-time slope, the oscillatory behaviour, and the long-time agreement with the analytical solution. Following Figure \ref{fig:convergence}, we use 500 terms to obtain a reliable analytical reference solution.

\subsection{Results with zero initial heat-flux derivative}
\begin{figure}[!htb]
  \centering
  \begin{subfigure}[t]{0.49\linewidth}
    \centering
    \includegraphics[width=\linewidth]{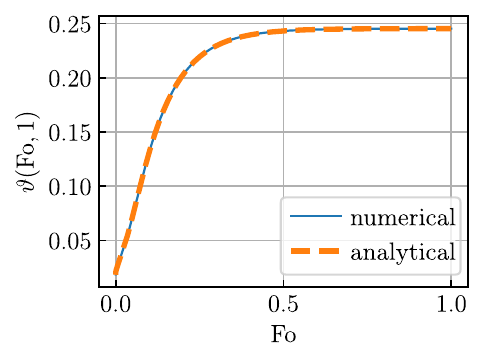}
    \caption{$\hat{\tau}=0.001$, temperature signal.}
  \end{subfigure}\hfill
  \begin{subfigure}[t]{0.49\linewidth}
    \centering
    \includegraphics[width=\linewidth]{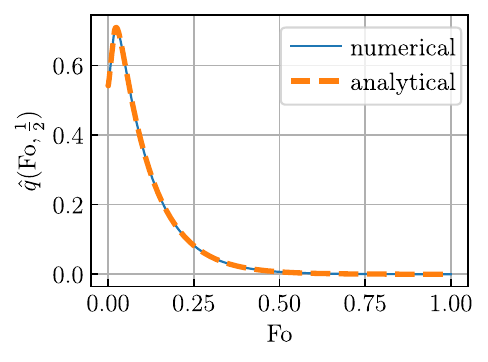}
    \caption{$\hat{\tau}=0.001$, heat-flux signal.}
  \end{subfigure}
  \begin{subfigure}[t]{0.49\linewidth}
    \centering
    \includegraphics[width=\linewidth]{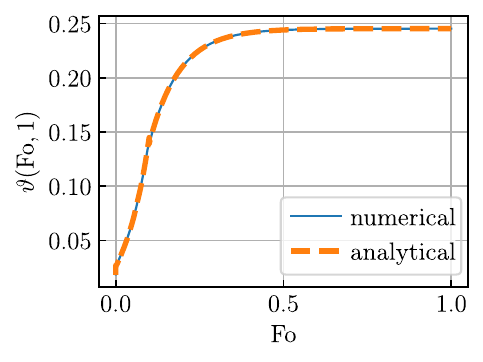}
    \caption{$\hat{\tau}=0.01$, temperature signal.}
  \end{subfigure}\hfill
  \begin{subfigure}[t]{0.49\linewidth}
    \centering
    \includegraphics[width=\linewidth]{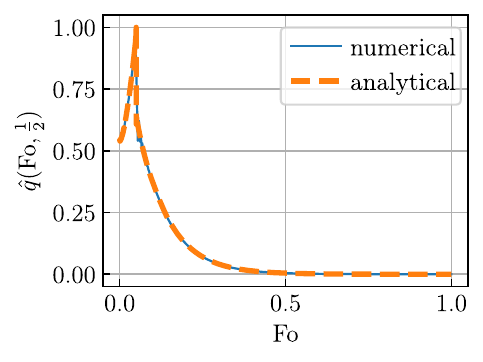}
    \caption{$\hat{\tau}=0.01$, heat-flux signal.}
  \end{subfigure}
  \begin{subfigure}[t]{0.49\linewidth}
    \centering
    \includegraphics[width=\linewidth]{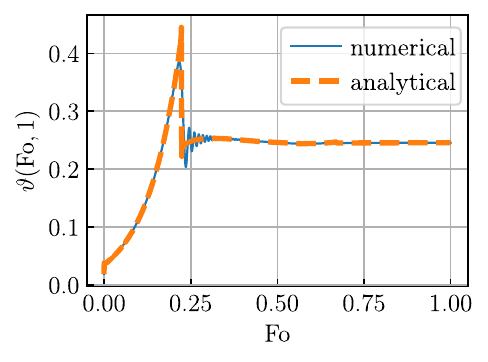}
    \caption{$\hat{\tau}=0.05$, temperature signal.}
  \end{subfigure}\hfill
  \begin{subfigure}[t]{0.49\linewidth}
    \centering
    \includegraphics[width=\linewidth]{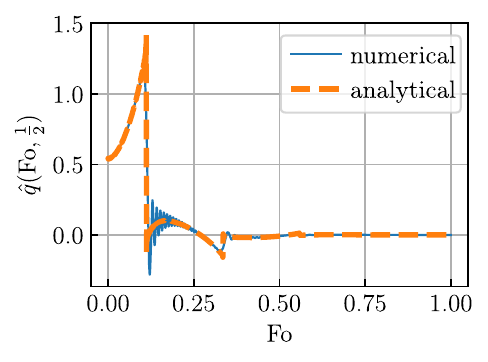}
    \caption{$\hat{\tau}=0.05$, heat-flux signal.}
  \end{subfigure}
  \caption{Comparison of analytical and numerical signals when the initial heat-flux derivative is set to zero.}
  \label{fig:mcv_article_zero}
\end{figure}

The first set of simulations corresponds to the simplest initialization, where the initial derivative is set to zero. Figure~\ref{fig:mcv_article_zero} summarizes the resulting temperature and heat-flux histories.
For $\hat{\tau}=0.001$, the agreement is acceptable despite the fact that it omits the dynamic behaviour, strictly speaking. At $\hat{\tau}=0.01$, the discrepancy becomes visible mainly in the heat-flux response, while the temperature curve still remains relatively close to the benchmark. At $\hat{\tau}=0.05$, the limitations of this approximation are clear; the numerically obtained heat flux deviates more strongly from the analytical solution, and oscillatory artifacts begin to appear. The iteration with this explicit scheme, therefore, captures the correct physical trend only in the weak-relaxation regime. 

\subsection{Results with a spatially uniform non-zero derivative}
\begin{figure}[!htb]
  \centering
  \begin{subfigure}[t]{0.49\linewidth}
    \centering
    \includegraphics[width=\linewidth]{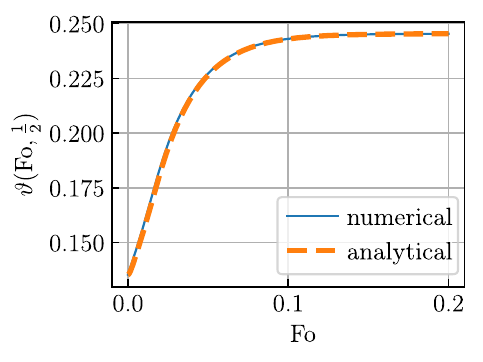}
    \caption{$\hat{\tau}=0.001$, temperature signal.}
  \end{subfigure}\hfill
  \begin{subfigure}[t]{0.49\linewidth}
    \centering
    \includegraphics[width=\linewidth]{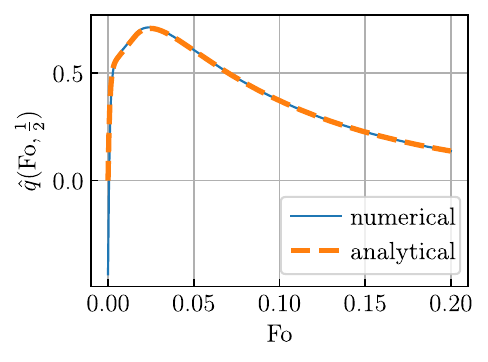}
    \caption{$\hat{\tau}=0.001$, heat-flux signal.}
  \end{subfigure}
  \begin{subfigure}[t]{0.49\linewidth}
    \centering
    \includegraphics[width=\linewidth]{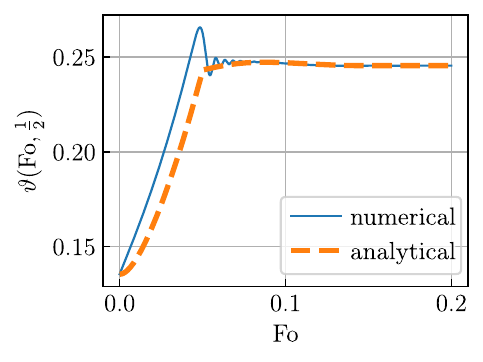}
    \caption{$\hat{\tau}=0.01$, temperature signal.}
  \end{subfigure}\hfill
  \begin{subfigure}[t]{0.49\linewidth}
    \centering
    \includegraphics[width=\linewidth]{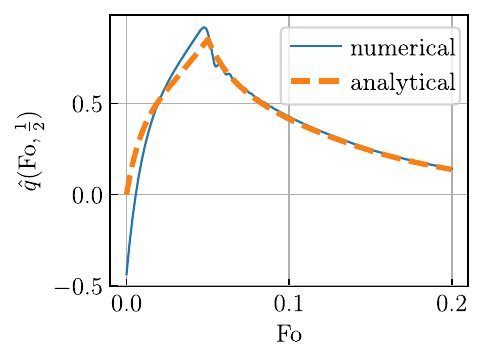}
    \caption{$\hat{\tau}=0.01$, heat-flux signal.}
  \end{subfigure}
  \begin{subfigure}[t]{0.49\linewidth}
    \centering
    \includegraphics[width=\linewidth]{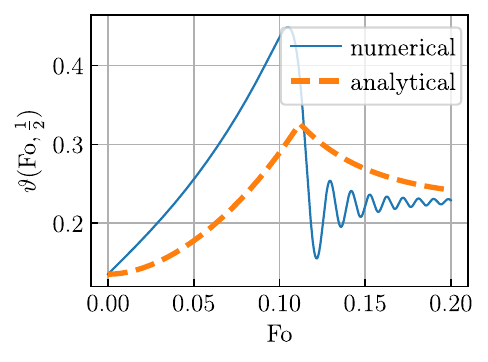}
    \caption{$\hat{\tau}=0.05$, temperature signal.}
  \end{subfigure}\hfill
  \begin{subfigure}[t]{0.49\linewidth}
    \centering
    \includegraphics[width=\linewidth]{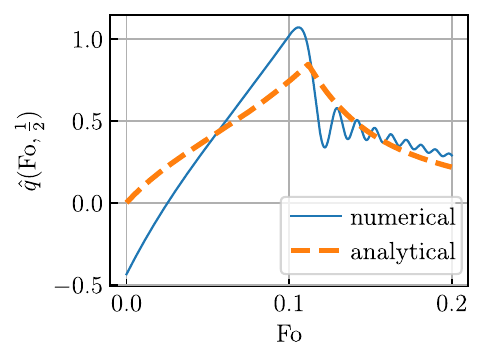}
    \caption{$\hat{\tau}=0.05$, heat-flux signal.}
  \end{subfigure}
  \caption{Comparison of analytical and numerical signals when a spatially uniform non-zero initial derivative of the heat flux is used.}
  \label{fig:mcv_article_const}
\end{figure}

The second set of simulations uses the constant derivative obtained from Eq.~\eqref{eq:mcv_article_qderiv_const}. The corresponding curves are collected in Figure~\ref{fig:mcv_article_const}.
This case is particularly insightful because it shows that an overly simplified implementation may lead to a poor numerical representation of the transient response. Compared with the zero-derivative case, the transient shape is modified, but the early-time behaviour is not captured correctly. In addition, strong non-physical oscillations appear in the numerical solutions for moderate and larger values of $\hat{\tau}$. Although this approximation is based on the assumption of a vanishing initial time derivative of the temperature field, its main advantage lies solely in its mathematical simplicity. Its accuracy is insufficient for reliable thermal evaluation or design-oriented use, and it does not improve by using a finer mesh or smaller time steps.

\subsection{Results with the space-dependent derivative}
\begin{figure}[!htb]
  \centering
  \begin{subfigure}[t]{0.49\linewidth}
    \centering
    \includegraphics[width=\linewidth]{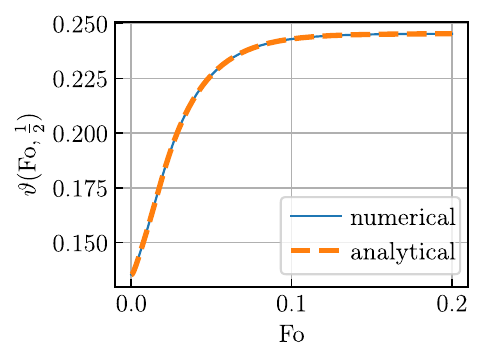}
    \caption{$\hat{\tau}=0.001$, temperature signal.}
  \end{subfigure}\hfill
  \begin{subfigure}[t]{0.49\linewidth}
    \centering
    \includegraphics[width=\linewidth]{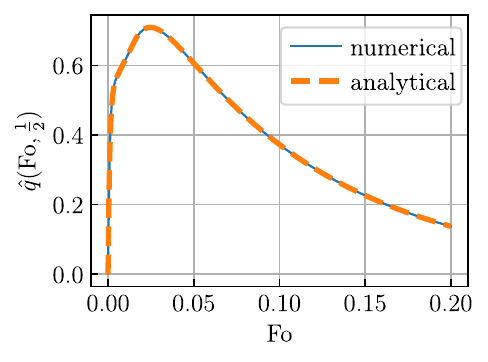}
    \caption{$\hat{\tau}=0.001$, heat-flux signal.}
  \end{subfigure}
  \begin{subfigure}[t]{0.49\linewidth}
    \centering
    \includegraphics[width=\linewidth]{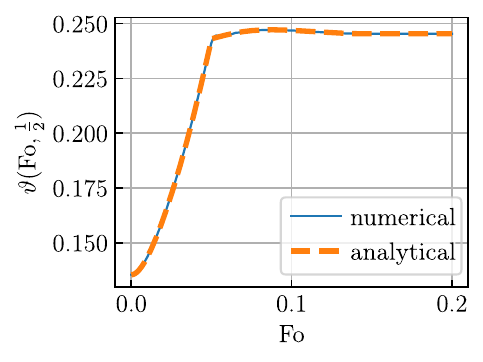}
    \caption{$\hat{\tau}=0.01$, temperature signal.}
  \end{subfigure}\hfill
  \begin{subfigure}[t]{0.49\linewidth}
    \centering
    \includegraphics[width=\linewidth]{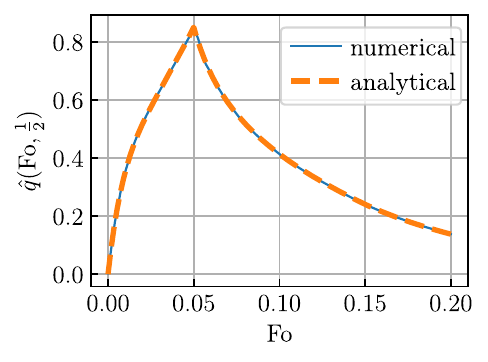}
    \caption{$\hat{\tau}=0.01$, heat-flux signal.}
  \end{subfigure}
  \begin{subfigure}[t]{0.49\linewidth}
    \centering
    \includegraphics[width=\linewidth]{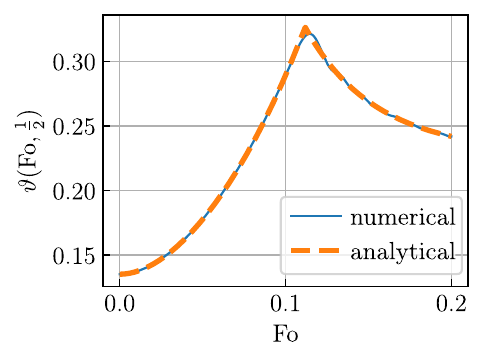}
    \caption{$\hat{\tau}=0.05$, temperature signal.}
  \end{subfigure}\hfill
  \begin{subfigure}[t]{0.49\linewidth}
    \centering
    \includegraphics[width=\linewidth]{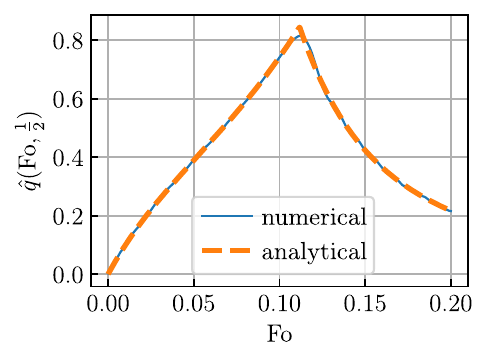}
    \caption{$\hat{\tau}=0.05$, heat-flux signal.}
  \end{subfigure}
  \caption{Comparison of analytical and numerical signals when the initial heat-flux derivative is implemented as a space-dependent field derived from the analytical reference.}
  \label{fig:mcv_article_field}
\end{figure}

The third set of simulations uses the full space-dependent derivative field given by Eq.~\eqref{eq:mcv_article_qderiv_pointwise}. The resulting temperature and heat-flux histories are shown in Figure~\ref{fig:mcv_article_field}.
This variant provides an excellent numerical solution under the assumption of a vanishing initial time derivative of the temperature field. In this case, the computed transient response agrees very well with the exact analytical reference, both for the temperature history and for the heat-flux evolution. The strong non-physical oscillations observed in the previous approximations do not appear. The quality of the agreement indicates that this mathematically rigorous initialization is highly suitable for design-oriented engineering applications. 

Furthermore, it is important to clarify that while this mathematical formulation is more demanding, the actual computational overhead is negligible. The modal summation is evaluated exactly once at $\mathrm{Fo}=0$, requiring $\mathcal{O}(N_x \times N)$ operations. This initial cost is mathematically insignificant compared to the explicit time-stepping loop that subsequently executes $\mathcal{O}(N_x \times N_t)$ times.

\begin{table}[!htb]
\centering
\small
\begin{tabularx}{\textwidth}{>{\raggedright\arraybackslash}p{2.9cm}|c|>{\centering\arraybackslash}p{2.3cm}|>{\centering\arraybackslash}p{2.1cm}|X}
Method & $\hat{\tau}$ & Temperature error [\%] & Heat-flux error [\%] & Main observation \\
\hline \hline
  & 0.001 & 0.0123 & 0.0329 & Excellent agreement in the near-diffusive regime. \\
Zero derivative & 0.01  & 0.0861 & 3.8639 & Visible deviation in the flux response. \\
  & 0.05  & 3.3973 & 18.3651 & Significant error growth and unphysical oscillations. \\
\hline
  & 0.001 & 0.1799 & 4.8930 & Changing the transient characteristics without improving accuracy. \\
Constant derivative & 0.01  & 2.1103 & 15.5982 & Errors become notable. \\
  & 0.05  & 14.5613 & 37.4268 & Strongly oscillatory and physically inaccurate solution. \\
\hline
& 0.001 & 0.0116 & 0.1315 & Deviates only slightly from the analytical solution. \\
Space-dependent derivative & 0.01  & 0.0334 & 0.4023 & Deviation remains small; calculation remains robust. \\
& 0.05  & 0.2171 & 1.2196 & Deviation increases moderately; physically accurate response.
\end{tabularx}
\caption{Quantitative comparison of the three initialization strategies against the analytical solution.}
\label{tab:mcv_article_comp}
\end{table}

\subsection{Quantitative error analysis}
For direct comparison, the relative error values obtained in the three cases are collected in Table~\ref{tab:mcv_article_comp}. The numerical error was evaluated as a discrete relative error in time. To capture the most representative dynamic behaviour, the temperature was monitored at the rear face ($i_\vartheta$ corresponding to $\xi=1$) and the heat flux was monitored at the midpoint of the domain ($i_q$ corresponding to $\xi=0.5$). The numerical data were compared with the corresponding values of the exact Galerkin reference analytical solution at the same time instants. The temperature error was computed as
\begin{equation}\label{eq:mcv_article_temp_error}
    \varepsilon_\vartheta
    =
    \frac{
    \sqrt{
    \displaystyle\sum_{j=0}^{N_t}
    \left(
    \vartheta_j^{\,i_\vartheta}
    -
    \vartheta_{\mathrm{ref},j}^{\,i_\vartheta}
    \right)^2
    }
    }{
    \sqrt{
    \displaystyle\sum_{j=0}^{N_t}
    \left(
    \vartheta_{\mathrm{ref},j}^{\,i_\vartheta}
    \right)^2
    }
    },
\end{equation}
while the heat-flux error was defined by
\begin{equation}\label{eq:mcv_article_flux_error}
    \varepsilon_{\hat{q}}
    =
    \frac{
    \sqrt{
    \displaystyle\sum_{j=0}^{N_t}
    \left(
    \hat{q}_j^{\,i_q}
    -
    \hat{q}_{\mathrm{ref},j}^{\,i_q}
    \right)^2
    }
    }{
    \sqrt{
    \displaystyle\sum_{j=0}^{N_t}
    \left(
    \hat{q}_{\mathrm{ref},j}^{\,i_q}
    \right)^2
    }
    }.
\end{equation}
Here, $\vartheta_j^{\,i_\vartheta}$ and $\hat{q}_j^{\,i_q}$ denote the numerical values at the selected spatial indices, while $\vartheta_{\mathrm{ref},j}^{\,i_\vartheta}$ and $\hat{q}_{\mathrm{ref},j}^{\,i_q}$ denote the corresponding exact Galerkin reference values.

From a practical engineering perspective, the discrepancies generated by improper initialization can be significant. As seen in Table~\ref{tab:mcv_article_comp}, utilizing a zero or uniform derivative at $\hat{\tau}=0.05$ yields an 18\% to 37\% error in the predicted heat flux. In high-frequency applications, an artificial 37\% overshoot or unphysical oscillation in the heat-flux prediction would lead to notable misinterpretations of localized thermal effects. For example, during biological ablation, it can be decisive in distinguishing between unintended tissue damage and the intended damage to cancer cells. Accurate initialization is therefore not just a mathematical formality, but a physical and practical necessity in modern engineering applications.

\section{Conclusion}\label{sec:mcv_conclusion_article}
In this work, the numerical initialization of the one-dimensional Maxwell--Cattaneo--Vernotte (MCV) heat conduction model was systematically investigated for an exponentially distributed initial temperature field and adiabatic boundary conditions. After introducing the dimensionless governing equations, an explicit finite-difference scheme was formulated on a staggered spatial grid, where temperature was assigned to cell centers and heat flux to cell faces. Because the MCV model contains the time derivative of the heat flux, the initial temperature distribution alone does not determine the first numerical step. Therefore, the main focus of this paper was the mathematically rigorous determination and discrete representation of this initial heat-flux derivative.

Three initialization strategies were compared with an analytically derived solution obtained via the Galerkin method. The zero heat-flux-derivative assumption yielded acceptable agreement only in the nearly diffusive regime. While not unphysical in itself, this assumption corresponds to a distinctly different initial non-equilibrium state than the one implied by a vanishing initial temperature derivative. As the dimensionless relaxation time $\hat{\tau}$ increased, the explicit finite-difference scheme became increasingly sensitive to this mismatch, and severe oscillatory deviations appeared in the computed response. The spatially uniform non-zero derivative introduced additional constitutive information; however, by compressing a spatially varying dynamic response into a single averaged value, it produced considerable deviations and step-like numerical artifacts for larger values of $\hat{\tau}$. 

The closest agreement with the analytical ground truth was achieved when the rigorously derived, full space-dependent derivative field was applied. This exact initialization robustly preserves the spatial thermodynamic structure associated with the selected analytical initial state. In the present explicit staggered scheme, both the temperature and heat-flux histories were kept practically identical to the reference solution over the entire investigated parameter range, without adding any significant computational runtime overhead.

Different physically meaningful initial assumptions lead to fundamentally different transient responses. As demonstrated, the numerical representation of these assumptions can heavily distort the observed solution—introducing spurious oscillations and dissipation errors—especially when relaxation effects are pronounced. Consequently, when evaluating extreme transient processes such as ultra-fast laser flashes on semi-transparent materials, biological tissue ablation, or thermal management in microelectronics, employing a mathematically rigorous initialization is critical. It is the only way to ensure that any observed wave-like phenomena reflect real physical material properties rather than numerical artifacts generated by the chosen time-integration scheme or commercial finite-element solvers.

A natural continuation of this work is the optimization of this procedure and its extension to the Guyer--Krumhansl (GK) model. Such an extension is particularly motivated by the fact that the GK model introduces higher-order spatial derivative terms to account for non-local interactions and over-diffusion \cite{Kovacs2024PhysRep}. Properly initializing these higher-order continuum models will introduce even greater mathematical complexities regarding the compatibility of boundary and initial conditions, making the systematic, fully coupled, and analytically grounded approach developed here an essential foundation for future investigations.

\section{Acknowledgement}
Project no.~TKP-6-6/PALY-2021 has been implemented with the support provided by the Ministry of Culture and Innovation of Hungary from the National Research, Development and Innovation Fund, financed under the TKP2021-NVA funding scheme. The research was funded by the Sustainable Development and Technologies National Programme of the Hungarian Academy of Sciences (FFT NP FTA). This work was partially supported in part by the Hungarian Scientific Research Fund under Grant agreement STARTING 149487.

\clearpage
\bibliographystyle{unsrt}
\bibliography{bibliography}

@article{Parker1961,
  title   = {Flash method of determining thermal diffusivity, heat capacity, and thermal conductivity},
  author  = {Parker, W. J. and Jenkins, R. J. and Butler, C. P. and Abbott, G. L.},
  journal = {Journal of Applied Physics},
  volume  = {32},
  number  = {9},
  pages   = {1679--1684},
  year    = {1961},
  publisher = {American Institute of Physics}
}

@article{Cattaneo1958,
  title   = {Sur une forme de l'equation de la chaleur eliminant la paradoxe d'une propagation instantanee},
  author  = {Cattaneo, C.},
  journal = {Comptes Rendus},
  volume  = {247},
  pages   = {431--433},
  year    = {1958}
}

@article{Vernotte1958,
  title   = {Les paradoxes de la theorie continue de l'equation de la chaleur},
  author  = {Vernotte, P.},
  journal = {Comptes Rendus},
  volume  = {246},
  pages   = {3154},
  year    = {1958}
}

@article{kovacs2022analytical,
  title   = {Analytical treatment of nonhomogeneous initial states for non-{F}ourier heat equations},
  author  = {Kov\'acs, R.},
  journal = {International Communications in Heat and Mass Transfer},
  volume  = {134},
  pages   = {106021},
  year    = {2022},
  publisher = {Elsevier}
}

@article{Tisza47, title={The theory of liquid {H}elium}, author={L. Tisza},  journal={Physical Review}, volume={72}, number={9}, pages={838-877}, year={1947}, publisher={APS}}

@ARTICLE{Lan47, Author = {L. Landau}, Title = {On the theory of superfluidity of {H}elium {II}}, Journal = {Journal of Physics}, volume={11},  number={1},  pages={91-92},  year={1947}}

@article{Kovacs2024PhysRep,
  title   = {Heat equations beyond {F}ourier: {F}rom heat waves to thermal metamaterials},
  author  = {Kov\'acs, R.},
  journal = {Physics Reports},
  volume  = {1048},
  pages   = {1--75},
  year    = {2024},
  publisher = {Elsevier}
}

@article{McN74t, title={Second {S}ound and {A}nharmonic {P}rocesses in {I}sotopically {P}ure {A}lkali-{H}alides},  author={McNelly, T. F.},  year={1974}, note={Ph.D. Thesis, Cornell University}}

@phdthesis{Feher2025PhD,
  author  = {Feh\'er, A. \'E.},
  title   = {Non-{F}ourier heat conduction in heterogeneous materials},
  school  = {Budapest University of Technology and Economics},
  year    = {2025},
  address = {Budapest}
}

@article{GK64, title={Dispersion relation for second sound in solids}, author={Guyer, R. A. and Krumhansl, J. A.},  journal={Physical Review}, volume={133}, number={5A}, pages={A1411}, year={1964}, publisher={APS}}

@book{MullerRuggeri1998,
  title     = {Rational {E}xtended {T}hermodynamics},
  author    = {Muller, I. and Ruggeri, T.},
  volume    = {37},
  year      = {2013},
  publisher = {Springer}
}

@article{VanFulop2012,
  title   = {Universality in heat conduction theory: weakly nonlocal thermodynamics},
  author  = {V\'an, P. and F\"ul\"op, T.},
  journal = {Annalen der Physik},
  volume  = {524},
  number  = {8},
  pages   = {470--478},
  year    = {2012},
  publisher = {Wiley}
}

@article{Auriault2016,
  title   = {Cattaneo--{V}ernotte equation versus {F}ourier thermoelastic hyperbolic heat equation},
  author  = {Auriault, J.-L.},
  journal = {International Journal of Engineering Science},
  volume  = {101},
  pages   = {45--49},
  year    = {2016},
  publisher = {Elsevier}
}

@book{ZolfMaer11b, title={Bioheat {T}ransfer}, author={Zolfaghari, A. and Maerefat, M.}, year={2011}, publisher={InTech}}

@article{JosephPreziosi1989,
  author  = {Joseph, D. D. and Preziosi, L.},
  title   = {Heat waves},
  journal = {Reviews of Modern Physics},
  year    = {1989},
  volume  = {61},
  number  = {1},
  pages   = {41--73},
  doi     = {10.1103/RevModPhys.61.41}
}

@article{OzisikTzou1994,
  author  = {{\"O}zisik, M. N. and Tzou, D. Y.},
  title   = {On the wave theory in heat conduction},
  journal = {Journal of Heat Transfer},
  year    = {1994},
  volume  = {116},
  number  = {3},
  pages   = {526--535},
  doi     = {10.1115/1.2910903}
}

@article{Chester1963,
  author  = {Chester, M.},
  title   = {Second sound in solids},
  journal = {Physical Review},
  year    = {1963},
  volume  = {131},
  number  = {5},
  pages   = {2013--2015},
  doi     = {10.1103/PhysRev.131.2013}
}

@article{Malina16,
  title={Ultrafast laser processing of materials: from science to industry},
  author={Malinauskas, M. and {\v{Z}}ukauskas, A. and Hasegawa, S. and Hayasaki, Y. and Mizeikis, V. and Buividas, R. and Juodkazis, S.},
  journal={Light: Science \& Applications},
  volume={5},
  number={8},
  pages={e16133--e16133},
  year={2016},
  publisher={Nature Publishing Group}
}

@article{Sharma20,
  title={Thermal management of {3-D} heterogeneously integrated microelectronics: challenges and future research directions},
  author={Sharma, M. K. and Ramos-Alvarado, B.},
  journal={Communications Engineering},
  volume={5},
  number={1},
  pages={28},
  year={2026},
  publisher={Nature Publishing Group UK London}
}

@article{Gu21,
  title={Material-structure-performance integrated laser-metal additive manufacturing},
  author={Gu, D. and Shi, X. and Poprawe, R. and Bourell, D. L. and Setchi, R. and Zhu, J.},
  journal={Science},
  volume={372},
  number={6545},
  pages={eabg1487},
  year={2021},
  publisher={American Association for the Advancement of Science}
}

@article{Walsh89,
  title={Er: {YAG} laser ablation of tissue: effect of pulse duration and tissue type on thermal damage},
  author={Walsh Jr, J. T and Flotte, T. J. and Deutsch, T. F.},
  journal={Lasers in Surgery and Medicine},
  volume={9},
  number={4},
  pages={314--326},
  year={1989},
  publisher={Wiley Online Library}
}

@article{Jau08,
  title={Bio-heat transfer analysis during short pulse laser irradiation of tissues},
  author={Jaunich, M. and Raje, S. and Kim, K. and Mitra, K. and Guo, Z.},
  journal={International Journal of Heat and Mass Transfer},
  volume={51},
  number={23-24},
  pages={5511--5521},
  year={2008},
  publisher={Elsevier}
}

@article{FehKov24,
title = {On the dynamic thermal conductivity and diffusivity observed in heat pulse experiments},
author = {Fehér, A. and Kovács, R.},
pages = {161--170},
volume = {49},
number = {2},
journal = {Journal of Non-Equilibrium Thermodynamics},
year = {2024},
}

@article{RietEtal18, title = {Implicit numerical schemes for generalized heat conduction equations}, journal = {International Journal of Heat and Mass Transfer}, volume = {126}, pages = {1177 - 1182}, year = {2018}, issn = {0017-9310}, author = {Rieth, Á. and Kovács, R. and Fülöp, T.}}

@article{FulEtal20, title={Thermodynamical extension of a symplectic numerical scheme with half space and time shifts demonstrated on rheological waves in solids}, author={Fülöp, T. and Kovács, R. and Szücs, M. and Fawaier, M.}, journal={Entropy}, volume={22}, pages={155}, year={2020}, publisher={MDPI}}

@article{Huang26,
  title={Heat conduction analysis of multi-directional {FGMs} with complex heat sources and boundary conductions using a {C}hebyshev spectral method},
  author={Huang, Y. and Zhao, Y. and Liu, H. and Xie, W. and Fei, M.},
  journal={Continuum Mechanics and Thermodynamics},
  volume={38},
  number={1},
  pages={3},
  year={2026},
  publisher={Springer}
}

@article{Amiri25I,
  title={A review on analytical heat transfer in functionally graded materials, {Part I}: {F}ourier heat conduction},
  author={Amiri Delouei, A. and Emamian, A. and Ghorbani, S. and Khorrami, A. and Jafarian, K. and Sajjadi, H. and Atashafrooz, M. and Jing, D. and Tarokh, A.},
  journal={Journal of Thermal Science},
  volume={34},
  number={4},
  pages={1358--1386},
  year={2025},
  publisher={Springer}
}

@article{Amiri25II,
  author    = {Amiri Delouei, A. and Emamian, A. and Ghorbani, S. and others},
  title     = {A {R}eview on {A}nalytical {H}eat {T}ransfer in {F}unctionally {G}raded {M}aterials, {Part II: Non-Fourier Heat Conduction}},
  journal   = {Journal of Thermal Science},
  volume    = {34},
  pages     = {1387--1407},
  year      = {2025}
}

\end{document}